\documentclass[lettersize,journal]{IEEEtran}

\usepackage{graphicx}
\usepackage{subfigure}
\usepackage{color}
\usepackage{epstopdf}
\usepackage{booktabs}

\usepackage{amsmath}
\allowdisplaybreaks

\usepackage{amssymb}
\usepackage{mathtools}
\usepackage{latexsym}
\usepackage{siunitx}

\usepackage{bm}
\usepackage{nicefrac}
\usepackage{booktabs}
\usepackage{array}
\usepackage{multirow}
\usepackage{threeparttable}
\usepackage{makecell}
\usepackage[procnumbered,ruled,vlined,linesnumbered]{algorithm2e}
\usepackage{stfloats}
\usepackage{graphicx}
\usepackage{color}
\usepackage{tabularx}
\usepackage{xcolor}
\usepackage{multirow}
\usepackage{makecell}
\usepackage{siunitx}
\usepackage{stfloats}
\usepackage{bm}
\usepackage{enumitem}
\usepackage{hyperref}

\def\G{\mathcal{G}}
\def\V{\mathcal{V}}
\def\E{\mathcal{E}}
\def\H{\mathcal{H}}

\newtheorem{theorem}{Theorem}[section]

\newtheorem{definition}[theorem]{Definition}

\newcommand\ones{\boldsymbol{\mathit{1}}}

\newcommand\xx{\boldsymbol{\mathit{x}}}

\newcommand\zz{\boldsymbol{\mathit{z}}}

\renewcommand\AA{\boldsymbol{\mathit{A}}}

\newcommand\DD{\boldsymbol{\mathit{D}}}
\newcommand\II{\boldsymbol{\mathit{I}}}

\newcommand\LL{\boldsymbol{\mathit{L}}}

\def\calG{\mathcal{G}}

\newcommand{\Hh}{{\cal H}}

\DontPrintSemicolon
\SetKw{KwAnd}{and}
\SetFuncSty{textsc}
\SetKwInOut{Input}{Input\ \ \ \ }
\SetKwInOut{Output}{Output}

\begin{document}

\title{Linear Opinion Dynamics Model with Higher-Order Interactions}
\author{Wanyue~Xu~\IEEEmembership{Student Member,~IEEE}, Zhongzhi~Zhang~\IEEEmembership{Member,~IEEE}

\thanks{This work was supported by the National Natural Science Foundation of China (No. 62372112  and No. U20B2051). (Corresponding author: Zhongzhi~Zhang.)\\ Wanyue Xu and Zhongzhi Zhang are with the Shanghai Key Laboratory of Intelligent Information Processing, School of Computer Science, Fudan University, Shanghai 200433, China.\\ E-mail: xuwy@fudan.edu.cn, zhangzz@fudan.edu.cn}}


\markboth{IEEE Transactions on Computational Social Systems}
{Xu \MakeLowercase{\textit{et al.}}: Linear Opinion Dynamics Model with Higher-Order Interactions}
\maketitle
\IEEEtitleabstractindextext{
\begin{abstract}
Opinion dynamics is a central subject of computational social science, and various models have been developed to understand the evolution and formulation of opinions. Existing models mainly focus on opinion dynamics on graphs that only capture pairwise interactions between agents. In this paper, we extend the popular Friedkin-Johnsen model for opinion dynamics on graphs to hypergraphs, which describe higher-order interactions occurring frequently on real networks, especially social networks. To achieve this, based on the fact that for linear dynamics the multi-way interactions can be reduced to effective pairwise node interactions, we propose a method to decode the group interactions encoded in hyperedges by undirected edges or directed edges in graphs. 
We then show that higher-order interactions play an important role in the opinion dynamics, since the overall steady-state expressed opinion and polarization differ greatly from those without group interactions. We also provide an interpretation of the equilibrium expressed opinion from the perspective of the spanning converging forest, based on which we design a fast sampling algorithm to approximately evaluate the overall opinion and opinion polarization on directed weighted graphs. Finally, we conduct experiments on  real-world hypergraph datasets, demonstrating the performance of our algorithm.

\end{abstract}

}


\IEEEdisplaynontitleabstractindextext
\IEEEpeerreviewmaketitle
\section{Introduction}
\IEEEPARstart{T}he widespread use of online social networks and social media~\cite{Le20} over the past few decades has constituted an integral part of people's lives~\cite{SmCh08}, which has significantly altered how people  exchange viewpoints and shape opinions~\cite{HaRaSaChKrHe22} on social platforms like  Facebook and Twitter. As a hot research subject of computational social science~\cite{HoWaAtetal21}, opinion dynamics has received considerable attention from the scientific community~\cite{BoCoDe20,ZhZh23}. There are numerous recent literature studying online opinion diffusion, evolution, and formation~\cite{DaGoMu14,VeGh21}. 
In real scenarios, users tend to communicate with like-minded individuals or familiar neighbors, leading to different sociological phenomena such as opinion polarization~\cite{MaTeTs17,MuMuTs18} and disagreement~\cite{MuMuTs18,Cho18}. In the current digital age, online social networks reinforce these phenomena and may create echo-chambers and filter bubbles~\cite{GiYuSa18}. However, the formation and influence mechanisms for polarization are still not well understood. 

An important step for studying opinion
dynamics is probably developing a mathematical
model. A rich variety of models
have been proposed for modelling opinion dynamics, among which the Friedkin-Johnsen (FJ) model~\cite{FrJo90} is popular one.  This linear model is simple but succinct, since it  incorporates French's ``theory of social power''~\cite{Fr56} to sufficiently capture  complicated social behaviors. Opinion shaping is significantly affected by the interactions among the individuals.  Most previous models for opinion dynamics only capture pairwise interactions among agents described by a graph, neglecting those higher-order interactions taking place among more than two entities at a time. However, various recent work show that in many realistic scenarios, interactions among three and more entities are ubiquitous in real natural~\cite{WuOtBa03,GiPaCuIt15,ReNoScect17} and social~\cite{PaPeVa17} networks. A typical example involving higher-order interactions is the scientific collaboration networks. For a paper with three or more authors, the collaboration relationship among all authors is higher order. 


There are different ways to encode or describe  higher-order interactions in networks, including affiliation graphs~\cite{NeWaSt02,Fe81}, simplicial complexes~\cite{CoBi17, PeBa18}, hypergraphs~\cite{Be89}, among others. In this paper, we focus on hypergraphs, since they are a convenient mathematical structure, capturing in a faithful and natural way the organization and characteristics of the above mentioned higher-order interactions, especially in social systems. As a fundamental organization structure, higher-order interactions have a critical consequence on various dynamical processes. For example, epidemic spreading~\cite{dePeMo20}, random walks~\cite{ChRa19,CaBaCe20}, and percolation on hypergraphs have been studied, producing novel collective behaviors with the main reason lying in the presence of higher-order interactions. 


Except for the aforementioned dynamical processes, opinion dynamics on hypergraphs has also been addressed recently. Very recently, based on hypergraphs, some models for opinion dynamics have been studied, including the DeGroot model~\cite{NeMeLa20},  the Deffuant-Weisbuch (DW) model~\cite{ScHe22} and Hegselmann-Krause (HK) model~\cite{HiKuBr22}.  It was shown~\cite{NeMeLa20} that linear opinion dynamics on hypergraphs can always be reduced to dynamics on graphs with appropriate pairwise interactions, and that nonlinear dynamics are necessary for higher-order interactions not to be reducible to pairwise interactions. Generally, there are two different way to transform/project higher-order interactions in hypergraphs to pairwise interactions in ordinary clique graphs with distinct vertex weights. One is hyperedge-dependent vertex weights~\cite{ChRa19,HaAkPa20}, and other is and hyperedge-independent vertex weights~\cite{AgLiZe05,ZhHuSc06,TaMiIk20}. The projection graphs can be either undirected or directed, depending on whether the pairwise relationships are symmetric or not~\cite{ZhHuSc05}. Although it is suggested that different pairwise interactions in projection graphs play a nonnegligible role in opinion dynamics, relevant study is still lacking. This motivates us to explore the effects of higher-order interactions on  linear opinion dynamics, concentrating on  FJ model.

Specifically, in this paper, we present a study of FJ opinion dynamics on hypergraphs, concentrating on the class of hypergraphs with hyperedge-dependent node weights. In such a hypergraph, there is a weight $h(e)$ for each hyperedge $e$ and a set of weights for every node $v$ with each weight $\gamma_e(v)$ representing the role of node $v$ in hyperedge $e$ incident to $v$. This kind of hypergraphs well describes the higher-order interactions observed in a wide range of real systems. For example, in  collaboration networks, a paper corresponds to a hyperedge $e$, and $\gamma_e(v)$ is associated the contribution of the author $v$, since generally the authors of every paper make different contributions. In the aspect of group chats in online social media, some individuals (e.g., administrators) in one group might be gregarious, presenting their opinions very frequently and affecting others' positions greatly, while other individuals might be reticent, having little impact on others. Such group relationship can also be modeled by hypergraphs with hyperedge-dependent node weights. In addition to social networks, hypergraphs with hyperedge-dependent node weights have also been applied to many other fields, such as image segmentation~\cite{DiYi10} and search~\cite{HuLiZh10}, text ranking~\cite{BeAl13}, and 3D object classification~\cite{ZhLiGa18}.

\begin{figure}
	\begin{center}
		\includegraphics[width=0.9\linewidth]{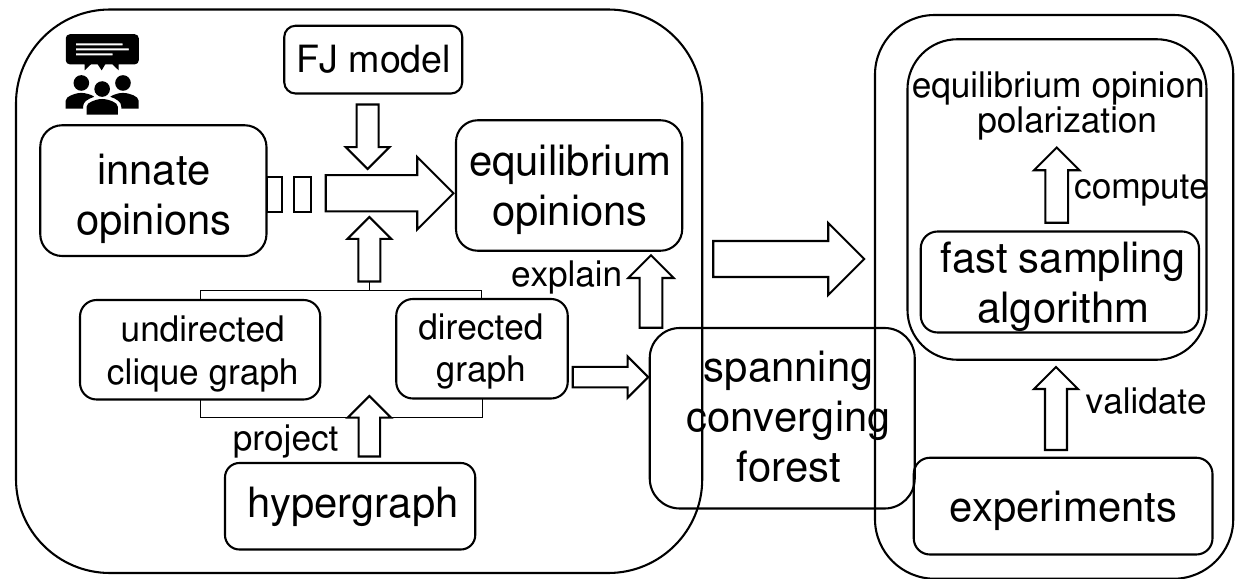}
		\caption{Graphical framework of this work.}
		\label{framework}
	\end{center}
\end{figure}

The framework of this work is shown in Figure~\ref{framework}.  
Our main contributions of this  paper are  as follows.
\begin{itemize}[leftmargin=*]
\item We propose an approach projecting a hypergraph into a weighted directed graph, which incorporates the higher-order organization and heterogeneous contributions of nodes to hyperedges in the hypergraph into the weights and directions of arcs in the corresponding digraph. 
	
\item We show that higher-order interactions have a substantial effect on opinion dynamics in a toy model hypergraph and real hypergraphs, focusing on the overall opinion and opinion polarization. 
	
\item We provide an interpretation of the expressed opinion in terms of spanning converging forests in the projected directed graph. 

\item We develop a fast sampling algorithm to approximately evaluate the overall opinion and opinion polarization in linear time. 

\item We perform extensive experiments on real hypergraph datasets, which demonstrate that our algorithm achieves both good efficiency and effectiveness, and scales to large networks. 

	
	
\end{itemize}

\section{Preliminaries}

In this section, we give a brief introduction to some essential concepts about graphs, hypergraphs, some related matrices such as  Laplacian, as well as the Friedkin-Johnsen opinion model on graphs.
\subsection{Graph}

Let $\G=(\V,\E, w)$ denote a weighted directed graph (often called digraph) with $n=|\V|$ nodes and  $m=|\E|$ arcs, where $\V$ is the node set $\{v_1,v_2,...,v_n\}$ and $\E$ is the arc set is  $\{e_1,e_2,...,e_m\}$. In the following text, $v_i$ and $i$ are used interchangeably to denote node $v_i$ if inducing no confusion. We use $i\to j$ to denote an arc in $\E$ pointing from $i$ to $j$. A self-arc is an arc having identical end nodes. An isolated node is a node having no arcs pointing to or coming from it. For two nodes $v_1$ and $v_k$ in digraph $\G$, a path from $v_1$ to $v_k$  is an alternating sequence of nodes and arcs $v_1, e_1, v_2,..., e_{k-1}, v_k$, in which nodes are distinct and every arc $e_i$ is $v_{i}\to v_{i+1}$. A loop is a path plus an arc from the ending node to the starting node. A graph is (strongly) connected if there is a path for any pair of nodes. A graph is called weakly connected if it is connected when one replaces any directed arc $i\to j$ with two directed arcs $i\to j$ and $j\to i$ in opposite directions. A tree is a weakly connected graph with no loops. An isolated node is considered as a tree. A forest is a particular graph that is a disjoint union of trees.

For a weighted digraph $\G$, many properties are encoded in its  weighted adjacency matrix $\AA$ of order $n \times n$, 
whose entry $w_{ij}$ at row $i$ and column $j$ represents the weight of arc $i\to j $. For each arc $i\to j \in \E$ in $\G$, its weight $w_{i j}$ is strictly nonnegative.
If there is no arcs from $i$ to $j$, $w_{i j}=0$. For a node $i$, its weighted in-degree $d^+_i$ is defined as  $d^+_i=\sum_{j=1}^n w_{ji}$, and its weighted out-degree $d^-_i$ is defined as $d^-_i=\sum_{j=1}^n w_{ij}$. In the sequel, we use $d_i$ to represent the out-degree $d_i^-$. For digraph $\G$, its out-degree diagonal matrix $\DD$ is an $n \times n$ matrix with the $i$th diagonal entry being $d_i$. The Laplacian matrix of $\G$ is $\LL=\DD-\AA$.
Let $\mathbf{1}$ and $\mathbf{0}$ be the two  $n$-dimensional vectors with all entries being ones and zeros, respectively.
Then, by definition, the sum of all entries in each row of $\LL$ equals $0$, namely,  $\LL\mathbf{1}=\mathbf{0}$.

For a weighted digraph $\G$, if for any arc $i\to j $ with weight $w_{ij}$, the arc
$ j \to i $ exists and has weight $w_{ji}$ equal to $w_{ij}$, then $\G$ reduces to an undirected weighted graph, denoted by $G$. Thus, for an  undirected graph $G$, $w_{ij}= w_{ji}$ holds for any pair of nodes $i$ and $j$, and $d^+_i= d^-_i $ holds for any node $i$. Moreover, both the weighted adjacency matrix $\AA$ and Laplacian matrix $\LL$ of $G$ are symmetric, satisfying $\LL\mathbf{1}=\mathbf{0}$.

\subsection{Hypergraph}

Let $\mathcal{H} = (\V,\textit{E}, h)$ be a weighted hypergraph with node set $\V$ and hyperedge set $\textit{E}\subset 2^{\V}$.  Each hyperedge $e \in \textit{E}$ is a subset of node set $\V$, representing the higher-order interactions among a group of nodes. Let $h(e)$ denote the weight of  hyperedge $e$, representing the   importance  of $e$ in $\mathcal{H}$. A hypergraph $\mathcal{H}$ is reduced a traditional graph $G$ when each hyperedge in $\mathcal{H}$ includes exactly two nodes, indicating a binary interaction between the two nodes. 

\begin{figure}
	\begin{center}
		\includegraphics[width=0.6\linewidth]{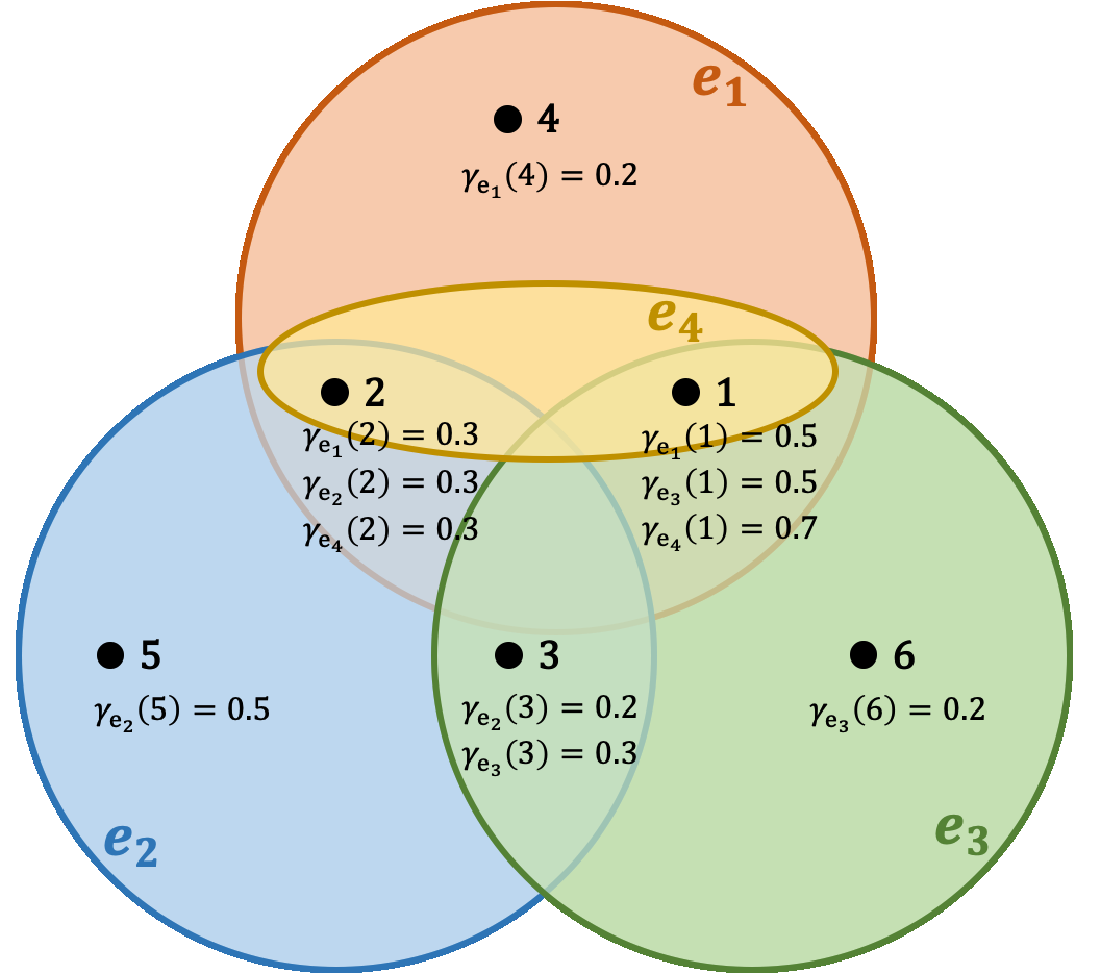}
		\caption{An example of a hypergraph with six nodes and four hyperedges $e_1$, $e_2$, $e_3$, $e_4$.}
		\label{hyper.eps}
	\end{center}
\end{figure}

For a hyperedge in a hypergraph, the contributions of its members are either inhomogeneous or homogeneous. Let $\gamma_{e}(i)$ denote the contribution proportion of node $i$ in the hyperedge $e$ including $i$. Then, for each hyperedge $e$, there is a collection of contribution fractions for its member nodes $\{\gamma_{e}(i_1), \gamma_{e}(i_2),..., \gamma_{e}(i_{|e|})\}$  satisfying $\sum_{i_k\in e}\gamma_{e}(i_k)=1$,  where $i_1$, $i_2$,..., $i_{|e|}$ belong to $e$. For two nodes $i$ and $j$ in $e$, if their contributions in $e$ are heterogeneous, $\gamma_{e}(i)\neq \gamma_{e}(j)$; if the contributions of all nodes in $e$ are homogeneous, $\gamma_{e}(i)= \gamma_{e}(j)=\frac{1}{|e|}$. Thus, for a node $i$ in hyperedge $e$, the contribution of $i$ to $e$ is $h(e)\,\gamma_{e}(i)$. For example, in collaboration networks, a hyperedge $e$ represents a paper, $h(e)$ denotes the impact of the paper, $\gamma_{e}(i)$ denotes the contribution faction of author $i$ to the paper, and $h(e)\,\gamma_{e}(i)$ represents $i$'s contribution to the whole paper. Figure~\ref{hyper.eps} illustrates a toy hypergraph with inhomogeneous contributions of nodes to each hyperedge.

\subsection{Friedkin-Johnsen Model on Graphs}
For the FJ opinion dynamics on a digraph $\G=(\V,\E, w)$, each node $i\in\V$ is associated with two opinions: one is the internal opinion $x_i$, the other is the expressed opinion $x_i^{(k)}$ at time $k \geq 1$. During the opinion evolution, the internal opinion remains unchanged,
while the expressed opinion $x_i^{(k)}$ evolves with time according to some given rules. In the classical FJ model, at time $k+1$, the expressed opinion  $x_i^{(k+1)}$ for agent $i$ is updated according to both of its internal opinion $x_i$ and its neighbors' expressed opinions at time $k$. Let $N(i)$ be the set of out-neighbors of node $i$, which are pointed by out-edges from $i$. 
Then, $x_i^{(k+1)}$ is a convex combination of  $x_{i}$ and $x_j^{(k)}$ with $j\in N(i)$ as
\begin{equation}\label{FJmodel}
x_i^{(k+1)}=\frac{x_{i}+\sum_{j\in N(i)}w_{ij}x_j^{(k)}}{1+\sum_{j\in N(i)}w_{ij}}.
\end{equation}

In~\eqref{FJmodel}, as a popular choice in the literature, we assume that  for all $i \in \V$, the weight of internal opinion is unit, and that the internal opinion $x_i$  is in the interval $[0,1]$.  The set of all internal opinions is denoted by vector  $\mathbf{x}=(x_1,x_2,\ldots,x_n) ^\top $ and the set of all expressed opinions at time $k$ is denoted by vector $\mathbf{x}^{(k)}= (x^{(k)}_1,x^{(k)}_2,\ldots,x^{(k)}_n) ^\top $. It was proved~\cite{BiKlOr15} that the expressed opinions evolving based on~\eqref{FJmodel} converge to a unique equilibrium opinion vector. Let $\mathbf{z}=(z_1, z_2,\ldots,z_n)^\top$ be vector of expressed opinions at steady state,  where  $z_i$ is the expressed opinion of node $i$. It was shown~\cite{BiKlOr15} that  $\mathbf{z}$  is the solution to the following linear system of equations when time $k$ tends to infinity:
\begin{equation}\label{steadystate}
\mathbf{z}=\mathbf{x}^{(\infty)}=(\mathbf{I}+\LL)^{-1}\mathbf{x},
\end{equation}
where  $\II$ is the $n$-dimensional identity matrix. 

Equation~\eqref{steadystate} shows that the equilibrium expressed opinion $\mathbf{z}$ is a linear combination of the internal opinions. Let $\omega_{ij}$ be the entry at row $i$ and column $j$ of matrix $\mathbf{\Omega}\triangleq\left(\II+\LL\right)^{-1}$, which is called the fundamental matrix of opinion dynamics~\cite{GiTeTs13}. Then, for every node $i \in \V$, its expressed opinion $z_i$ is given by  $z_i=\sum^n_{j=1}  \omega_{ij}x_j$. Thus, the equilibrium expressed opinion for every node is determined by the internal opinions, as well as the interactions between nodes encoded in matrix $\LL$.

\section{Opinion Dynamics on Hypergraphs}
In existing works on the FJ model for opinion dynamics, only the pairwise interactions are taken into account, ignoring those  group interactions that frequently occur in social media and social  networks.  
In this section, we generalize the FJ model on traditional graphs to hypergraphs, which are a popular mathematical paradigm describing higher-order interactions. 
{{It was proved that for linear dynamics models,  higher-order interactions in hypergraphs can be converted to pairwise interactions  in corresponding effective graphs~\cite{NeMeLa20}.}} 

\subsection{Projection of Hypergraphs}
In prior works,  projection methods were used to capture the higher-order interactions  on hypergraphs, in order to uncover their influences on various dynamical processes such as epidemic spreading~\cite{dePeMo20} and random walks~\cite{ChRa19,CaBaCe20} as well as other related problems about 
 hygergpraphs, such as clustering~\cite{HaAkPa20} and embedding~\cite{ZhHuSc07}. 

{\textbf{Projection to undirected clique graphs.}} In existing works~\cite{AgLiZe05,ZhHuSc06,TaMiIk20}, the higher-order interactions of a hypergraph $\mathcal{H}=(\V,E,h)$ are studied by mapping the nodes belonging to a hyperedge into a weighted clique of appropriate size in a projected clique graph $H=(\V,E',h)$. The graph $H$ is undirected with the same node set $\V$ as $\H$. The set $E'$ of edges in $H$ is defined by $E' = \{(i, j) | i\in e, j \in e \textrm{ for some } e \in E, i \neq j\}$. Thus, each hyperedge $e\in E$ in $\H$ is transformed to a clique of $|e|$ nodes and $|e|(|e|-1)/2$ edges in the corresponding projected graph $H$. For each edge with end nodes $i$ and $j$, its weight $w_{ij}$ is defined by 
\begin{equation}\label{wij1}
w_{ij}=\sum_{\substack{i\in e,j\in e,\\ e \in E, i\neq j}} 2\left(h(e)\times \frac{1}{|e|}\times \frac{1}{|e|-1}\right).
\end{equation}

\begin{figure}
	\begin{center}
		\includegraphics[width=0.9\linewidth]{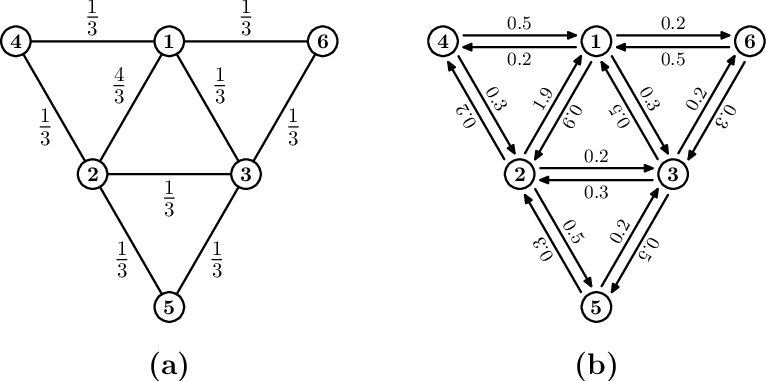}
		\caption{Illustration of the projected clique graph (a) and the projected directed graph (b) for the hypergraph shown in Fig.~\ref{hyper.eps}.}
		\label{projected.eps}
	\end{center}
\end{figure}

In this way, a hypergraph $\mathcal{H}=(\V,E,h)$
is transformed  into a projected clique graph $H=(\V,E',h)$. Note that when hypergraph $\mathcal{H}$ is a traditional graph, the weight between nodes $i$ and $j$ defined in~\eqref{wij1} for the projected graph  is consistent with the edge weight in the original graph. In the  projected clique graph, the weights of all edges in a clique corresponding to a hyperedge are identical, equaling the weight $2h(e)$ divided by $|e|(|e|-1)$. Figure~\ref{projected.eps}(a) shows the projected clique graph of the hypergraph illustrated in Figure~\ref{hyper.eps}. 

{\textbf{Projection to directed graphs.}} As shown in the  preceding section, in many practical scenarios, the role of each member in a hyperedge is heterogeneous, which is neglected in the projected undirected clique graphs. To incorporate this heterogeneity of higher-order structure, we propose another mapping approach that projects hypergraph $\mathcal{H}=(\V,E, h)$ into a weighted digraph $\mathcal{G}=(\V,\E, h)$. The projected digraph $\mathcal{G}$ has the same node set as the original hypergraph $\mathcal{H}$. The arc set $\E$ in $\G$ is defined as $\E=\{(i \to j) | i\in e, j \in e \textrm{ for some } e \in E, i \neq j\}$.
Thus, each hyperedge $e$ in $\mathcal{H}$ corresponds to $|e|(|e|-1)$ directed arcs in $\mathcal{G}$, connecting $|e|$ nodes belonging to hyperedge $e$. For each arc $ i \to j $ in $\mathcal{G}$, its weight is defined by 	
\begin{equation}\label{wij2}
w_{ij}=\sum_{\substack{i\in e,j\in e,\\ e \in E, i\neq j}} 2\left(h(e)\times\gamma_{e}(j)\times \frac{1}{|e|-1}\right).
\end{equation} 
We note that~\eqref{wij1} is a particular case of~\eqref{wij2}, since \eqref{wij2} is reduced to~\eqref{wij1} when 
$\gamma_e(j)= \frac{1}{|e|}$. Figure~\ref{projected.eps}(b) illustrates the projected directed graph of the hypergraph  in Fig.~\ref{hyper.eps}. 

\subsection{The FJ Model on Hypergraphs}
After projecting hypergraph $\mathcal{H}$ to digraph $\mathcal{G}$, opinion dynamics on $\mathcal{H}$ is reduced to that on $\mathcal{G}$, with the higher-order interactions in $\mathcal{H}$ being encoded in the directions and weights of arcs in $\mathcal{G}$. Thus, similarly to the standard FJ model on digraphs shown in~\eqref{FJmodel}, in the FJ model on hypergraphs, at each time step, every agent updates its expressed opinion according to its internal opinions and the current expressed opinions of its neighbors in the same hyperedges. According to~\eqref{wij2}, the weights $w_{ij}$ and $w_{ji}$ for the pair of arcs $i\to j$ and $j \to i$ in the projected directed graph $\G$ are usually unequal, except that $\gamma_e(i)= \gamma_e(j)$ for every hyperedge $e$ containing $i$ and $j$. This asymmetric interaction is stemmed from the heterogeneous contribution of each member in a hyperedge, and is ubiquitous in social systems.

At first glance, one might hypothesize that the interaction between node pairs defined in~\eqref{wij2} does not capture higher-order organizations in hypergraphs, since $w_{ij}$ is defined based on a directed graph containing edges only between node pairs. Below we will show that $w_{ij}$ given in~\eqref{wij2} indeed grasps the higher-order interactions on the corresponding hypergraph, at least for opinion dynamics. For this purpose, we define a probability transition matrix $\mathcal{P}$ associated with a random walk on a hypergraph $\mathcal{H} = (\V,E, h)$, with the $i$th diagonal entry being $p_{ii}=1/({1+\sum_{j\in N(i)}w_{ij}})$ and the $ij$th non-diagonal entry defined by $p_{ij}=w_{ij}/({1+\sum_{j\in N(i)}w_{ij}})$. Then the updating rule for opinion dynamics in $\mathcal{H}$ defined in~\eqref{FJmodel} can be rewritten as 
\begin{equation}\label{FJmodelp}
x_i^{(k+1)}=p_{ii}x_{i}+ \sum_{j\in N(i)} p_{ij} x_j^{(k)}.
\end{equation}


In the case that hypergraph $\mathcal{H}=(\V,E,h)$ is transformed to a projected clique graph $H=(\V,E',h)$, the interaction between nodes $i$ and $j$ is described by~\eqref{wij1}. The probability transition matrix $\mathcal{P}$ does not encode higher-order relations between nodes, since the transition probabilities $p_{ij}$ are fully determined by the weights of edges in the undirected projected clique graph $H$, as shown by Theorem 3.1 in~\cite{ChRa19}. Therefore, to define the transition probability between two nodes $i$ and $j$ for random walks on $\mathcal{H}$, only the information of a single quantity $w_{ij}$ is required, which encodes a pairwise relation between $i$ and $j$. As such, the probability transition matrix $\mathcal{P}$ for random walks on $\mathcal{H}$ is equivalent to that of the undirected graph $H$, which indicates that $\mathcal{P}$ only captures pairwise relations between nodes in $\mathcal{H}$. 

By contrast, when the hypergraph $\mathcal{H}=(\V,E,h)$ is projected to a weighted digraph $\mathcal{G}=(\V,\E, h)$, the node interaction is represented by~\eqref{wij2}. In this case, the probability transition matrix $\mathcal{P}$ can encode the higher-order interactions between nodes in $\mathcal{H}$, since the transition probability $p_{ij}$ cannot be derived only from the quantity $w_{ij}$ defined for the pair of nodes $i$ and $j$, for reasons see Theorem 3.2 in~\cite{ChRa19}.  The higher-order asymmetric interactions between agents in opinion dynamics produce new collective behaviors of opinions as will be shown later.

\section{Interpretation of Steady-State Opinions}\label{interpret}
After reducing the FJ model on a hypergraph $\mathcal{H}$ to the FJ model on an associated directed graph $\mathcal{G}$, we next provide an explanation for the equilibrium opinions in terms of the spanning converging forests in  $\mathcal{G}$.

For a directed graph $\mathcal{G}=(\V,\E,w)$, a subdigraph of $\G$ is another digraph, in which the sets of the nodes and arcs are subsets of $\V$ and $\E$, respectively, but the node set must include all endpoints of all arcs. A spanning subdigraph of $\G$ is a subdigraph with the same node set $\V$. A spanning converging tree or an in-tree is a weakly connected digraph, in which one node named the root or the sink node has out-degree $0$ and any other node has out-degree $1$. An isolated node is considered as a spanning converging tree with the root being itself. A spanning converging forest of $\G$ is a spanning subdigraph of $\G$, where all weakly connected components are converging trees. A spanning converging forest is also called an in-forest~\cite{AgCh01,ChAg02}. 

\begin{figure}
	\begin{center}
		\includegraphics[width=0.9\linewidth]{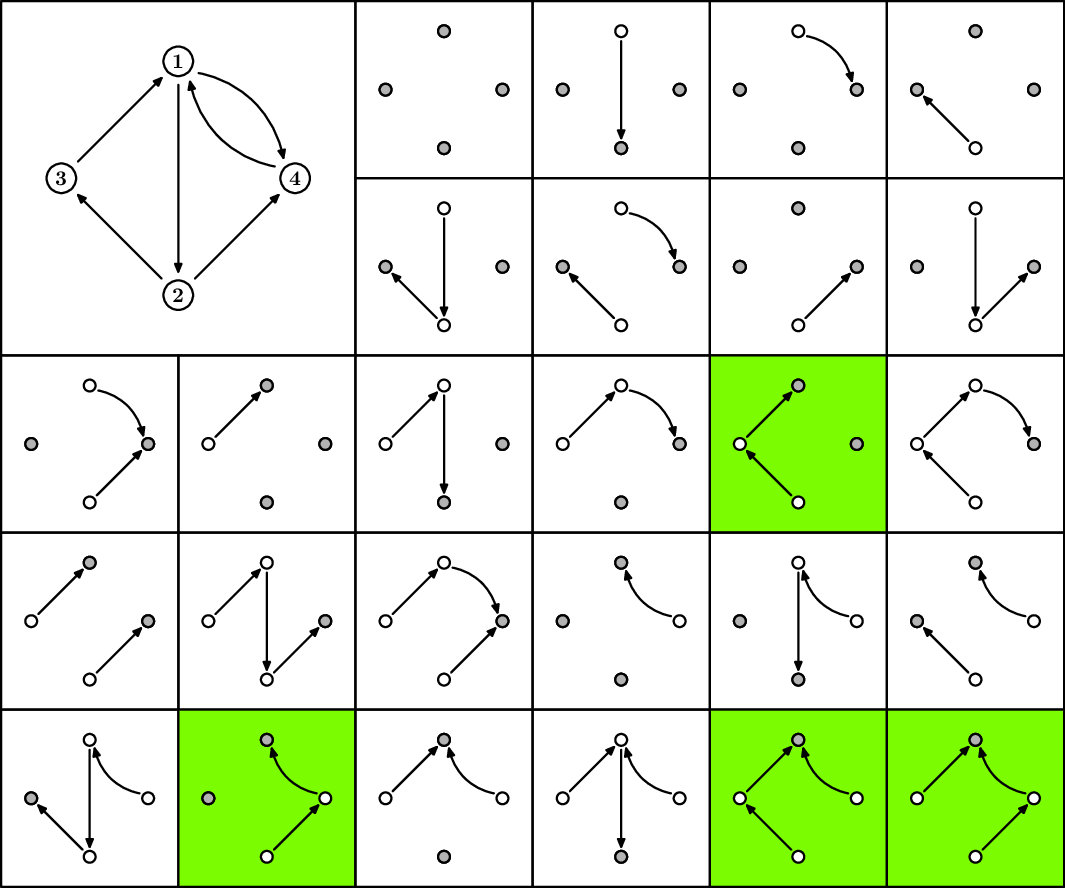}
		\caption{A digraph and  its  in-forests. The 4 in-forests with green background constitute the set $\Gamma_{21}$, for each in-forest in which $v_1$ and $v_2$ belong to the same in-tree with $v_1$ being the sink node.  
		}
		\label{G0}
	\end{center}
\end{figure}

For a subdigraph $\G'$ of digraph $\G$, its weight $\varepsilon(\G')$ is defined as the product of the weights of all arcs in $\G'$. For the case that  $\G'$ has not any arc, the  weight $\varepsilon(\G')$ is set to be $1$. Let $S$ be a nonempty set  of subdigraphs, define the weight of  $S$ as $\varepsilon(S) =\sum_{\G' \in S}  \varepsilon(\G')$. If $S$ is an  empty set, its weight is defined to be zero~\cite{ChSh97,ChSh98}. For a digraph $\G$, let $\Gamma$ be the set of all its in-forests, and $\Gamma_{i j}$ the set of in-forests with nodes $i$ and $j$ in the same in-tree rooted at node $j$. If every edge in $\G$ has unit weight, then $ \varepsilon(\Gamma)$ is equal to the total number of in-forests of $\G$, and $\varepsilon(\Gamma_{i j}) $ equals the number of spanning rooted forests of $\G$, where nodes $i$ and $j$ are in the same tree rooted at $j$. For example, for the graph with 4 nodes and 6 arcs on the top left corner of Figure~\ref{G0}, it has exactly $26$ in-forests, among which there are 4 in-forests where $v_2$ belongs to a converging tree sinked in $v_1$, that is $\varepsilon(\Gamma)=26$ and $\varepsilon(\Gamma_{21})=4$.  

For the particular case when $\G$ is an undirected weighted graph $G$, a spanning forest on $G$ is a spanning subgraph of $G$, which is a forest. A spanning forest of $G$ is called a spanning rooted forest, if each tree in the forest has a node marked as its root. For a subgraph $G'$ of $G$, its weight $\epsilon(G')$ is defined as the product of the weights of all undirected edges in $G'$. If $G'$ is an empty graph, its weight is set to be 1. For any nonempty set $S$ of subgraphs, its weight $\epsilon(S)$ is defined by $\epsilon(S) =\sum_{ G' \in S} \epsilon(G')$. For an empty $S$, its weight is set to be $0$. For each spanning rooted forest of $G$, one can construct a unique in-forest in the following way: for every tree in the spanning rooted forest, assign a direction for its edges so that its root is still the sink of the directed tree, that is, the root node is reachable from any other node in the directed tree.

In~\cite{ChSh97,ChSh98}, the  in-forest matrix of a digraph $\G$ is introduced and defined  as  $\mathbf{\Omega}=\left(\II+\LL\right)^{-1}$, where the entry $\omega_{ij}$ at row $i$ and column $j$ is determined by the weights $\varepsilon(\Gamma_{i j})$ and $\varepsilon(\Gamma)$, given by  $\omega_{ij}=\varepsilon(\Gamma_{i j})/ \varepsilon(\Gamma)$. Obviously, the in-forest matrix $\mathbf{\Omega}$ is exactly the fundamental matrix of opinion dynamics. Matrix $\mathbf{\Omega}$ has many good properties. For example, $\omega_{ij}\geq 0$ holds for any pair of nodes $i$ and $j$, with  $\omega_{ij}= 0$ if and only if there is no path  from   $i$ to $j$~\cite{Me97}. 

By definition of $\mathbf{\Omega}$ and  Figure~\ref{G0}, the entry  $\omega_{21}$ of in-forest matrix $\mathbf{\Omega}$ corresponding to the graph on the left top corner is $\varepsilon(\Gamma_{21})/ \varepsilon(\Gamma)=4/26$. Moreover, other entries of the in-forest  matrix $\mathbf{\Omega}$ can also be determined analogously as given by
\begin{align}
\mathbf{\Omega}=(\II+\LL)^{-1}=\frac{1}{26}\left(\begin{array}{cccc}12 & 4 & 2 & 8 \\ 4 & 10 & 5 & 7 \\ 6 & 2 & 14 & 4 \\ 6 & 2 & 1 & 17\end{array}\right).
\end{align}

 As shown above, the equilibrium expressed opinion $z_i$ for any node $i \in \V$  is  $z_i=\sum^n_{j=1}  \omega_{ij}x_j$. Here we provide an  explanation to the factor $\omega_{ij}$ in terms of the weights of in-forests, which is the ratio of  $\varepsilon(\Gamma_{i j})$ to  $\varepsilon(\Gamma)$.

Note that in~\cite{GiTeTs13,AbKlPaTs18}, $\omega_{ij}$ was interpreted as the probability of a random walker starting from node $i$ to be  absorbed in the copy node of $j$. In~\cite{BiKlOr15}, the expressed opinion was explained in terms of game theory. In addition, when the graph is unweighted, the expressed opinion can be also accounted for from the perspective of electrical networks~\cite{GhSr14}. Our  interpretation in terms of in-forest matrix is novel, different from those previous explanations.

\section{Effects of Higher-Order Interactions on Opinions}\label{interaction}

In this subsection, we study
the influence of higher-order interactions on opinion dynamics base on the FJ model, focusing on the
overall expressed opinion and the opinion polarization.

\subsection{Difference between Fundamental Matrices}
As shown above, for a given vector of internal opinions the equilibrium opinions are determined by the fundamental matrix $\mathbf{\Omega}= (\II+ \LL)^{-1}$. To uncover the influences of higher-order interactions, we distinguish two cases of higher-order interactions, described by~\eqref{wij1} and~\eqref{wij2}, respectively.  

For the first case, the interactions between any pair of nodes are equal to each other. In this case, the Laplacian matrix $\LL$ is symmetric, it is the same with the fundamental matrix $\mathbf{\Omega}= (\II+ \LL)^{-1}$. Moreover, $\mathbf{\Omega}$ is a doubly stochastic matrix~\cite{Me97}, which means that any entry of $\mathbf{\Omega}$ is nonnegative,  with each row and column summing up to $1$.

For the second case of higher-order interactions deciphered in~\eqref{wij2}, the interactions $w_{ij}$ and $w_{ji}$ between a pair of nodes $i$ and $j$ are unequal to each other, since the Laplacian matrix $
\LL$ is probably asymmetric. Although the fundamental matrix $\mathbf{\Omega} = (\II + \LL)^{-1} $ is always row stochastic, it might not be  column stochastic.

Let us consider two projected graphs in Figure~\ref{projected.eps} for the hypergraph in Figure~\ref{hyper.eps}. Although both graphs have six nodes and seemingly similar structure, they differ in several aspects. For example,  Figure~\ref{projected.eps}(a) is undirected, while  Figure~\ref{projected.eps}(b) is directed. In addition, the in-forest matrix of the graph in~Figure~\ref{projected.eps}(a) is a doubly stochastic matrix; while the  in-forest matrix of the graph in~Figure~\ref{projected.eps}(b) is row stochastic but not column stochastic, since $\mathbf{\Omega}\ones = \ones$, and $\ones^{\top} \mathbf{\Omega}= (1.4714, 0.8388, 0.8392, 0.8122, 1.2260, 0.8123)$.

\subsection{Effects on Overall Opinion}

A fundamental  quantity for opinion dynamics is the overall expressed opinion,  defined as the sum of expressed opinions over all nodes, which has received much recent attention~\cite{GiTeTs13,AbKlPaTs18,ZhZhLiZh23}, due to its practical  applications. For example,  favorable overall opinion about a specific information item is a key quantity in promotion campaigns and recommendation systems~\cite{GiTeTs13}.

By~\eqref{steadystate}, the expressed opinion of node $i\in \V$ is $z_i=\sum_{j=1}^n
\omega_{ij}x_j$. Then, the overall expressed opinion is $\sum_{i=1}^n z_i=\sum_{i=1}^n\sum_{j=1}^n
\omega_{ij}x_j$. For doubly stochastic $\mathbf{\Omega}$, $\sum_{i=1}^n
\sum_{j=1}^n \omega_{ij}x_j=\sum_{j=1}^n\sum_{i=1}^n \omega_{ij}x_j$$=\sum_{j=1}^n
x_j$. Thus, the overall expressed opinion is equivalent to the sum of internal opinions of all nodes, although the expressed opinion at equilibrium for a single agent might differ from its internal opinion. For directed graphs, the asymmetry of $\mathbf{\Omega}$   leads to the disappearance of conservation of total opinions, that is, $\sum_{i=1}^n z_i\neq \sum_{i=1}^n x_i$. Thus, the higher-order interactions between agents have substantial influences on opinion dynamics, especially the overall expressed opinion.

For the hypergraph in Figure~\ref{hyper.eps}, suppose that the internal opinion vector for both projected graphs is identical, equaling $\mathbf{x}=(0.1, 0.2, 0.3, 0.4, 0.5,  0.6)^{\top}$. We use \eqref{steadystate} to derive the equilibrium expressed opinion.
For the projected undirected clique graph, the overall internal opinion and the overall equilibrium expressed opinion are both 2.1. The equilibrium expressed opinion vector is $(0.2540, 0.2714, 0.3319, 0.3451, 0.4206, 0.4772)^{\top}$. While for the projected directed graph, the equilibrium expressed opinion vector is (0.2222, 0.2536, 0.3159, 0.3262, 0.4262, 0.4477$)^{\top}$, the  overall equilibrium expressed opinion is $1.992$, less than the overall internal opinion $2.1$. If we change the internal opinion vector to (0.6, 0.5, 0.4, 0.3, 0.2, 0.1$)^{\top}$ with the same overall opinion as $\mathbf{x}$, the expressed opinion vector in the projected clique graph equals (0.4461, 0.4287, 0.3682, 0.35501, 0.2794, 0.2229$)^{\top}$, with  the overall equilibrium opinion being 2.1. In contrast, the equilibrium expressed opinion vector in the projected directed graph is (0.4778, 0.4464, 0.3841, 0.3738, 0.2738, 0.2523$)^{\top}$, indicating that the overall equilibrium expressed opinion is 2.2080, larger than 2.1. Thus, the higher-order interactions that are incorporated into asymmetric arcs can not only increase but also decrease the  overall expressed opinion, which is the optimization objective of many recent works on opinion dynamics~\cite{GiTeTs13,AbKlPaTs18,ZhZhLiZh23}.

\subsection{Effects on Opinion Polarization}

The enormous popularity of social media and online social networks generates or strengthens some social phenomena, for example, polarization. The classical FJ model of opinion dynamics seems to be a good model to study related social phenomenon phenomena such as polarization~\cite{MuMuTs18, XuBaZh21,ZhBaZh21,XuZhGuZhZh22}. In the FJ model, if the equilibrium expressed opinions have an increased divergence, we say that opinion formation dynamics is polarizing. Intuitively, polarization measures the deviation degree of the equilibrium expressed opinions for all nodes from their average. Different indicators have been proposed to quantify polarization. As in many existing works, we adopt the metric proposed  in~\cite{MuMuTs18} to measure polarization.
\begin{definition}
	For a graph  $\G= (\V,\E,w)$,  let $ \bar{z} $ be the mean-centered equilibrium vector given by
	$ \bar{\zz} =z  - \frac{\zz^{\top}  \textbf{1}}{n} \textbf{1}$. Then the opinion polarization $P(\G)$ is defined to be:
	\begin{equation}\label{eq:dfn_contr}
	P(\G) =  \sum\limits_{i \in V}\bar{z} _i^2 =\bar{\zz}^{\top} \bar{\zz}.
	\end{equation}
\end{definition}

Polarization is affected by various factors, such as network structure~\cite{MuMuTs18}, filter bubbles~\cite{ChMu20},   and stubbornness~\cite{XuZhGuZhZh22}.  Below we explore the effects of higher-order interactions on opinion polarization.

For both of the above two vectors of internal opinions in the  hypergraph in Figure~\ref{hyper.eps}, the opinion polarization is 0.0369 in the its projected undirected clique graph, and  is 0.0407 in the corresponding  projected directed graph. Thus, the opinion polarization formed in the projected directed graph is larger than that in the undirected clique graph. The reason for this strengthening in the projected directed graph lies in the higher-order interactions captured by the  weight size and the direction of arcs. Our result is consistent with  the conclusion in~\cite{TaMiIk20}  that  groups often exert influence on individual members through the discussion of decision alternatives.  In projected directed graph, due to the asymmetric relations encoded in the directed edges, the probability of the appearance of extreme  opinions increases under the effects of higher-order interactions. 


In addition to the model hypergraph in Figure~\ref{hyper.eps}, higher-order interactions  also significantly affect the overall opinion and polarization in real-world hypergraphs, as we will show in Experiments Section.

\section{Fast Sampling Algorithm}

By~\eqref{steadystate}, direct computation of the overall expressed opinion and opinion polarization involves inverting the matrix $\II + \LL$, which takes $O(n^3)$ time and is  thus computationally impractical for large graphs. In fact, even if all the entries of the fundamental matrix $\mathbf{\Omega}$ are  known in advance, computing the overall expressed opinion and opinion polarization by~\eqref{steadystate} takes $O(n^2)$ time, which is still unacceptable for large networks with one million nodes. Here we design a linear-time  approximate algorithm by sampling spanning converging forests. 

   
In~\cite{AvLuGaAl18,PiAmBaTr21}, an extension of Wilson's algorithm~\cite{Wi96} was proposed to generate a rooted  spanning converging forest  $\phi_0$ from $\Gamma$. This algorithm is applicable to any digraph $\mathcal{G}=(\V,\E, h)$, connected or disconnected. Besides the Wilson's algorithm, another ingredient of the extended algorithm is the loop-erasure operation on a random walk first introduced in~\cite{La80}. For any spanning converging forest $\phi \in \Gamma $, the extended Wilson algorithm returns $\phi$ with probability $\mathbb{P}(\phi_0 =\phi)$ proportional to the weight of $\phi$.  In other words, for any forest $\phi $ in $\Gamma$, we have 
$\mathbb{P}(\phi_0 =\phi) = \frac{\varepsilon(\phi)}{\varepsilon({\Gamma})}$,
which is the theoretical justification of our fast algorithm.  

In order to enable the extended Wilson's algorithm to estimate the overall expressed opinion and opinion polarization, we rewrite the expressed opinion $z_i$ of node $i$ as
\begin{equation}\label{rf}
\begin{aligned}
    z_i &= \sum_{j=1}^n \omega_{ij}x_j = \sum_{j=1}^n \frac{\varepsilon(\Gamma_{ij})}{\varepsilon(\Gamma)}x_j \\ &=  \sum_{j=1}^n \sum_{\phi\in \Gamma_{ij}}\frac{\varepsilon(\phi)}{\varepsilon(\Gamma)}x_j = \sum_{j=1}^n \mathbb{P}(\phi \in \Gamma_{ij}) x_j.
\end{aligned}
\end{equation}
Leveraging~\eqref{rf} and the extended Wilson's algorithm~\cite{AvLuGaAl18,PiAmBaTr21}, we present a fast sampling algorithm to estimate the vector $\mathbf{z}$ for the equilibrium expressed opinions  and the opinion polarization $P(\calG)$.  The detailed of our algorithm is presented in Algorithm~\ref{alg-rf}.

\begin{algorithm}
	\caption{$\textsc{Sample}(\calG, \mathbf{x}, \tau$)}
	\label{alg-rf}
	\Input{ $\calG$ : a digraph, $\mathbf{x}$: internal opinion vector,   $\tau$: number of samples\\
	}
	\Output{$\widehat{\mathbf{z}}$:  equilibrium  expressed opinion, $\widehat{P}(\calG)$: opinion polarization }
	Inforest[$ i $] $ \leftarrow $ false,	Nextnode[$ i $] $ \leftarrow -1$, Rootindex[$i$] $ \leftarrow 0 $, $\widehat{\mathbf{z}}$[$i$] $ \leftarrow 0 $, $i= 1,2,\ldots,n $\;
            \For{$ j = 1 $ to $ \tau $}{
\For{$ i = 1 $ to $ n $}
	{$ u \leftarrow i $\; 
		\While{not Inforest[$ u $]}{
			seed $ \leftarrow $ \textsc{Rand}()\quad \% a random seed from $(0,1)$  \; 
			\If{seed  $\leq  \frac{1}{1+d_u} $}{
				Inforest[$ u $] $ \leftarrow $ true\;
				Nextnode[$ u $]$ \leftarrow -1 $\;
				Rootindex[$ u $] $ \leftarrow u $\;
			}
			\Else{
				Next[$ u $] $ \leftarrow $ \textsc{RandomSuccessor}($ u,\calG $)\; 
				u $ \leftarrow $ Next[$ u $]\;
			}
		}
		Rootnow $ \leftarrow $ 	Rootindex[$ u $], $ u\leftarrow i $\;
		\While{not Inforest[$ u $]}{
			Inforest[$ u $] $ \leftarrow $ true\;
			Rootindex[$ u $] $ \leftarrow $ Rootnow\;
			u $ \leftarrow $ Nextnode[$ u $]\;
		}
	}
	\For{$ i = 1 $ to $ n $}{
                $u$ $\leftarrow$ Rootindex[$i$]\;
            $\widehat{\mathbf{z}}[i]$ $\leftarrow$ $ \widehat{\mathbf{z}}[i] $+ $x_u/\tau$
    }
            }
    $ \bar{\zz}$ $\leftarrow$ $\widehat{z}- \frac{\widehat{z}^{\top}\textbf{1}}{n} \textbf{1}$\;
        $\widehat{P}(\calG) \leftarrow \bar{\zz}^{\top} \bar{\zz} $\\
	\textbf{return} $\widehat{\mathbf{z}}, \widehat{P}(\calG)$\;
\end{algorithm}

Below we show that for any node $i$, the $i$-th entry  $\widehat{z}_i$ of vector $\widehat{\mathbf{z}}$ returned by Algorithm~\ref{alg-rf} is an unbiased estimator of the equilibrium expressed opinion $z_i$. For convenience  of expression, let ${\rm I}\{\}$ be an indicator function, which takes the value $1$ if the statement inside the brackets is true, and $0$ otherwise. Algorithm~\ref{alg-rf} generates $\tau$ spanning converging forests $\phi_1, \phi_2,\ldots, \phi_{\tau}$, and computes   $\widehat{z}_i$ by the following expression 
\begin{equation}
    \widehat{z}_i = \frac{1}{{\tau}}{\sum_{j=1}^n\sum_{u=1}^{\tau} {\rm I}\{\phi_u \in\Gamma_{ij} \} x_j}.
\end{equation}
Then, the expectation of $\widehat{z}_i$ is 
\begin{align}
    \mathbb{E}(\widehat{z}_i) =& \mathbb{E}\Big(\frac{1}{\tau}\sum_{j=1}^n\sum_{u=1}^{\tau} {\rm I}\{\phi_u \in\Gamma_{ij} \} x_j \Big) \nonumber\\
    =&\sum_{j=1}^n \mathbb{P}(\phi \in \Gamma_{ij}) x_j = z_i,
\end{align}
which shows that $\widehat{z}_i$ is an unbiased estimator of  $z_i$ for every node $i$. 

Finally, based on an analysis similar to that in~\cite{PiAmBaTr21}, we obtain that the time complexity of Algorithm~\ref{alg-rf} is $O(\tau(n+m))$, which is linear with the sum of the number $n$ of nodes  and  the number $m$ of edges in the projected directed graph $\calG$.

\begin{table*}[]
\Large
\caption{Statistics of Hypergraphs,  their projected directed graphs, and projected clique graphs, as well as results for overall opinion and  polarization on projected graphs obtained by algorithms $\textsc{Sample}$ and \textsc{Exact}.}
\label{table}
\resizebox{\linewidth}{!}{
\begin{tabular}{lccccccccccc}
\hline
\multicolumn{1}{c}{{hypergraphs}} & nodes & {hyperedges} & {\begin{tabular}[c]{@{}c@{}}directed\\ edges\end{tabular}} & {\begin{tabular}[c]{@{}c@{}}undirected\\ edges\end{tabular}} & {sum($\mathbf{\xx}$)} &{sum($\zz$)} & {sum($ \widehat{\zz}$)} & \textbf{$P_1$} & \textbf{$P_2$} & \textbf{$P_3$} & \textbf{$P_4$} \\ \hline
email-Enron & 143 & 10883 & 3628 & 2681 & 68.87 & 72.07(+4.46\%) & 72.67 & 0.016 & 0.051 & 0.017 & 0.052 \\
contact-high-school & 327 & 172035 & 11636 & 5937 & 157.78 & 145.09(-8.04\%) & 144.28 & 0.001 & 0.017 & 0.002 & 0.02 \\
tags-stack-overflow & 49998 & 14458875 & 14298946 & 8469924 & 24468.3232 & 24735.96(+1.09\%) & 24981.16 & 88.04 & 143.59 & 88.99 & 145.35 \\
coauth-MAG-Geology & 1256385 & 1590335 & 13798864 & 6934647 & 579248.48 & --- & 579318.16 & --- & --- & 43501.56 & 47148.62 \\
coauth-DBLP & 1924991 & 3700067 & 23906448 & 12289310 & 924903.01 & --- & 924838.31 & --- & --- & 55311.86 & 61030.41 \\
threads-stack-overflow & 2675955 & 11305343 & 45192908 & 22665445 & 1276467.61 & --- & 1277074.45 & --- & --- & 59979.96 & 65630.32 \\ \hline
\end{tabular}}
\end{table*}

\begin{table}[]
\tiny
\caption{Running time (seconds) of algorithms $\textsc{Sample}$ and \textsc{Exact} on hypergraph datasets.}
\label{table_time}
\resizebox{\linewidth}{!}{
\begin{tabular}{lcc}
\hline
\multicolumn{1}{c}{{hypergraphs}} & {\begin{tabular}[c]{@{}c@{}}running time \\ of \textsc{Exact}\end{tabular}} & {\begin{tabular}[c]{@{}c@{}}running time \\ of \textsc{Sample}\end{tabular}} \\ \hline
email-Enron & 0.0061 & 0.0171  \\
contact-high-school & 0.0061 & 0.091 \\
tags-stack-overflow & 1537.6237 & 63.2271  \\

coauth-MAG-Geology & -- & 210.6149 \\
coauth-DBLP & -- & 405.403 \\
threads-stack-overflow & -- & 605.2165 \\ \hline
\end{tabular}}
\end{table}

\section{Experiments}\label{experiment}
In this section, we evaluate our algorithm $\textsc{Sample}$ on both projected clique graphs and projected directed graphs corresponding to real-world hypergraph datasets, and compare the  results with those obtained by  \textsc{Exact} method via inverting matrix $\II+\LL$ as shown in~\eqref{steadystate}, in order to demonstrate the  effectiveness and efficiency of $\textsc{Sample}$. 

\subsection{Setup}

\textbf{Environment.} Both algorithms $\textsc{Sample}$ and \textsc{Exact} are written in Julia. All our experiments are performed on an x64 machine with E5-2690 
2.60GHz Intel Xeon CPU and 112GB of memory. The operation system is Ubuntu 20.04 LTS. 

\textbf{Datasets.} We assemble a diverse collection of  real-world datasets for hypergraphs in our experiments.  In each hypergraph, nodes and hyperedges have their practical meanings. For example, in coauthorship networks (e.g., coauth-DBLP), a hyperedge corresponds to a set of authors publishing an article. In email networks (e.g., email-Enron), nodes are email addresses and a hyperedge is comprised of the sender and all recipients of an email. In online social networks (e.g., threads-stack-overflow), nodes are users and users in a common hyperedge represent that they participate in the same topic. The selected real higher-order datasets are publicly available in \url{https://cs.cornell.edu/~arb/data/}.

\textbf{Projection and parameter settings.} We first project these real hypergraphs into undirected clique graphs and directed graphs by~\eqref{wij1} and~\eqref{wij2}, respectively. We neglect the hyperedges with only one node, since we only care about the nodes interacting with others. For simplicity, we set the weight of each hyperedge to be 1. It is sufficient to unreal the impact of unbalanced higher-order relationships among nodes in each hyperedge on opinion evolution. For the projected directed graphs, we assume that the weights of nodes in the same hyperedge follow a power-law distribution $x^{-2}$, and normalize them to interval $[0,1]$. This assignment strategy is designed to simulate the unevenness in the hierarchical system of real-world networks. For example, in a chat group of a social software, the weight of administrator is much higher than an ordinary member, since it is more influential, and vice versa. The number $\tau$ of samplings in algorithm \textsc{Sample} is set to be $1000$. For each node $i$, its internal opinion $x_i$ is uniformly generated in the range of $[0,1]$.

The statistics of hypergraphs and their corresponding projected graphs are listed in the first five columns of Table~\ref{table}. 


\subsection{Results}

\textbf{Effects of high-order interactions.} In Table~\ref{table}, we report our experiment results, where ${\rm sum}({\xx})$ and ${\rm sum}(\zz)$ represent the overall internal opinion and overall equilibrium expressed opinion on projected directed graphs, computed by \textsc{Exact}. Since the overall internal opinion and overall equilibrium expressed opinion on an undirected graph are identical, we do not list them in Table~\ref{table}. The percentages in brackets denote $({\rm sum}(\zz)- {\rm sum}({\xx}))/ {\rm sum}({\xx})$, that is the relative difference between ${\rm sum}({\xx})$ and ${\rm sum}(\zz)$. ${\rm sum}( \widehat{\zz})$  represents the results of the overall equilibrium opinion on digraphs obtained by \textsc{Sample}. $P_1$, $P_2$, $P_3$, and $P_4$ in the four rightmost columns of  Table~\ref{table} are, respectively, the results of opinion polarization on undirected graphs by \textsc{Exact}, directed graphs by \textsc{Exact}, undirected graphs by \textsc{Sample}, and directed graphs by \textsc{Sample}. 

Table~\ref{table} shows that the  difference ${\rm sum}(\zz)- {\rm sum}({\xx})$ can be positive or negative, which shows that higher-order interactions have a significant effect on the overall expressed opinion in all tested hypergraphs. In voting and election scenarios, which is an important application field of opinion dynamics, a few percent of overall opinion deviations will have a serious impact on the final result. Moreover, the quantity of polarization in every directed graphs is larger than that on its corresponding undirected graph, which implies that higher-order asymmetric interactions play a vital role in opinion polarization on hypergraphs. Thus,  the existence of  higher-order interactions in real-world social networks may lead to  increasing opinion polarization.


\textbf{Accuracy.} To demonstrate the accuracy of algorithm \textsc{Sample}, we compare the results of overall equilibrium expressed opinion and opinion polarization for  \textsc{Exact} and \textsc{Sample} and report the results in Table~\ref{table}. We can see that for each network,  \textsc{Sample} always returns a value close to the exact solution for \textsc{Exact}. For example, ${\rm sum}(\zz)$ is close to ${\rm sum}( \widehat{\zz})$, it is the same with $P_1$ and $P_2$ ($P_3$ and $P_4$). 


\textbf{Efficiency.}  Except for the high accuracy,  algorithm \textsc{Sample} is also efficient, compared with \textsc{Exact}. Table~\ref{table_time} reports the running time of the two algorithms \textsc{Sample} and \textsc{Exact} on different networks considered in the experiments. From Table~\ref{table_time}, we observe that for small networks with less than ten thousand nodes, the running time for  \textsc{Sample} is larger than that for \textsc{Exact}. While for moderately large networks with about tens of thousands of nodes, \textsc{Sample} is much faster than \textsc{Exact}. For example, for the tags-stack-overflow network with about 50,000 nodes, \textsc{Sample} returns the results in about one minute, while  \textsc{Exact} takes more than 30 minutes to return the results. With respect to very large networks, \textsc{Sample} is highly efficient. For the last three networks with over  100,000 nodes, algorithm \textsc{Exact} fails to run, due to the limitations of high memory and time cost, while \textsc{Sample}  still returns the results quickly.  Particularly, \textsc{Sample} is scalable to large networks with millions of nodes, such as threads-stack-overflow with more than two million nodes.

Therefore, algorithm \textsc{Sample} is both effective and efficient, and scales to massive graphs.

\section{Related Work}

In this section, we provide a brief review of previous work related to ours.



\textbf{Models for opinion dynamics.} 
Many relevant models for opinion dynamics have been presented~\cite{PrTe17}, among which the DeGroot model is probably the first model proposed by  DeGroot~\cite{De74} and French~\cite{Fr56}. 
A significant extension of the DeGroot model if the FJ model~\cite{FrJo90},  which grasps the real social behaviors by incorporating ``theory of social power''~\cite{Fr56}. Since its establishment, the FJ model has been extensively studied, including the sufficient condition for the stability~\cite{RaFrTeIs15}, the average innate opinion~\cite{DaGoPaSa13}, the unique equilibrium expressed opinion vector~\cite{DaGoPaSa13,BiKlOr15}, and explanations~\cite{GhSr14,BiKlOr15}. Diverse variants of the FJ model have also been introduced~\cite{JiMiFrBu15}, by incorporating different factors affecting opinion formation, such as susceptibility to persuasion~\cite{AbKlPaTs18}, peer-pressure~\cite{SeGrSqRa19}, and algorithmic filtering~\cite{ChMu20}.
Both the DeGroot model and FJ model are linear, and there are several continuous nonlinear models with the Hegselmann-Krause model~\cite{HeKr02} and the Deffuant-Weisbuch model~\cite{DeNeAm00} being two well-known examples that consider bounded confidence.

\textbf{Social phenomena.} Some social phenomena have been quantified and optimized  based on the FJ model~\cite{XuBaZh21}, such as disagreement~\cite{MuMuTs18}, conflict~\cite{ChLiDe18}, polarization~\cite{MaTeTs17,MuMuTs18}, and controversy~\cite{ChLiDe18}. In real settings, opinion can converge, polarize, or  fragment, depending on the processes of opinion dynamics~\cite{HeKr02}. Nowadays people can communicate  conveniently on social media and online social networks, which make opinion tend to polarize, rather than reach consensus~\cite{GiYuSa18}.   In~\cite{ChMu20}, the impact of filter bubbles on polarization was discussed by including  algorithmic filtering to the FJ model. Opinion optimization problems for the FJ model are also proposed and studied~\cite{GiTeTs13,AbKlPaTs18}. 


\textbf{Higher-order interactions.}  Higher-order interactions are  ubiquitous in real networked systems~\cite{ScLiRoLaStBe02}, particularly social networks~\cite{BrHuGe06,RhShHoLeKiCh11}. They belong to the class of group interactions taking place among three or more entities simultaneously, which can be modeled by hypergraphs~\cite{Be89}. The influences of higher-order interactions on various dynamics and functions based on hypergraphs have attracted considerable attention from the scientific community. In~\cite{ZhHuSc07}, the powerful methodology of spectral clustering on undirected graphs was generalized to hypergraphs, and some algorithms were developed for hypergraph embedding and transductive classification. In~\cite{ChRa19}, random walks on hypergraphs with edge-dependent vertex weights were studied, based on which a flexible framework for clustering hypergraph-structured data was proposed in~\cite{HaAkPa20}. In~\cite{ZhCuJiCh18},  the problem of hyperedge prediction was addressed in hypergraphs. For opinion dynamics on hypergraphs, most existing studies focus on non-linear opinion models such as DW model~\cite{ScHe22} and HK model~\cite{HiKuBr22}, giving too little care to the linear FJ opinion model.  This motivates us to explore the effects of higher-order interactions on FJ model.

\section{Conclusion}

In this paper, we extended the classic FJ  opinion dynamics model to hypergraphs, taking into account the higher-order interactions.  We resolved this kind of linear dynamics problem by pairwise network representation. To this end, we projected a hypergraph with homogeneous hyperedges to an undirected weighted graph, while transformed   a hypergraph with inhomogeneous hyperedges to a directed  weighted graph. We provided an explanation of the equilibrium expressed opinion in terms of spanning converging forests. We proved that  higher-order interactions have a non-negligible  influence on opinion dynamics. Moreover, we proposed a fast  algorithm by sampling spanning converging forest to approximately solve the problem concerned in linear time. Extensive experiments on real-life hypergraphs validate the performance of our algorithm. {{Further work should include applying our techniques and algorithms to  other dynamics on hypergraphs.

\bibliographystyle{IEEEtran}
\bibliography{hyper,internalopinion}

\end{document}
\endinput